\documentclass[twocolumn,preprintnumbers,amsmath,amssymb]{revtex4}
\usepackage{amsmath,amsfonts,latexsym,amssymb,graphics,epsfig,subfigure,color,makeidx}
\usepackage{xcolor}
\usepackage{multirow}
\newcommand {\nn}    {\nonumber}

\begin{document} \large

\title{Generalized  Nonsingular  Solutions  for  the Scalar Meson  Field of a  Point Charge  in General  Relativity Theory
\footnote{\large Translated by E. J. Saletan}
}

\author{Yi-Shi Duan }

\affiliation{\large{Moscow State  University}}

\begin{abstract}\large
\begin{center}
            (Submitted  to  JETP editor September 15, 1956)  \\
          J.~Exptl.~Theoret.~Phys.~(U.S.S.R.)~31,~1098-1099~\\
          (December,~1956)
\end{center}
\end{abstract}

\maketitle
It  has been shown that in investigating the electromagnetic and meson fields of elementary particles, the effects of  gravitational interaction cannot be neglected \cite{shirokov1,fisher2,shirokov3,duan4,enistein5}. The present note  is an attempt to find  solutions to the equations of the general relativistic gravitational and scalar meson fields of a  point nuclear charge. The solution is nonsingular at all points of the gravitational and meson fields.  Although
it contains several functions  whose  form  is  not given, this solution can be  used to obtain the  potential of the scalar meson field, which  is a  generalization of the Yukawa potential;  furthermore,  it makes  it possible to calculate  the  mass and self-energy of the nucleon, which turn out to be finite.

We  take the general centrally symmetric expression for the  interval in the  usual form
\begin{equation}
ds^{2}=-e^{\lambda}dr^2-e^{\mu}r^2(d\theta^2+\sin^{2}\theta d\Phi^2)+e^{\nu}dt^2,\nn
\end{equation}
where $\lambda,~\mu$ and $\nu$ are functions of $r$.  The energy-momentum tensor of the scalar meson field is
\begin{eqnarray}
T^{k}_{i}&=& \Big(\frac{1}{8\pi}\Big)
    \left\{2g^{ik}\frac{\partial U}{\partial x^{i}}\frac{\partial U}{\partial x^{k}} \right. \nonumber\\ 
    &-&\left. \delta^{k}_{i}\Big(g^{lm}\frac{\partial U}{\partial x^{l}}\frac{\partial U}{\partial x^{m}}-\chi^2U^2\Big)\right\}.\nn
\end{eqnarray}
We  shall consider the  field static and set $U=U(r)$. Then
\begin{eqnarray}
\begin{array}{l}
  T^{1}_{1}=\left(\frac{1}{8\pi}\right)\left(-e^{-\lambda}U'^{2}+\chi^2U^2\right); \\~\\
  T^{2}_{2}=T^{3}_{3}=T^{4}_{4}=\left(\frac{1}{8\pi}\right)\left(e^{-\lambda}U'^{2}+\chi^2U^2\right).
\end{array} \label{1}
\end{eqnarray}

It  is easy to show,  on the  basis of  the  invariance  of  the  Einstein equations under the transformation $r\rightarrow\sigma r,  \chi\rightarrow\chi/\sigma$, where $\sigma$ is a constant, that the solution  to these equations is of the form
\begin{eqnarray}
\begin{array}{l}
  U\,=U_{0}V(\chi r); ~~~e^{\nu}=e^{\nu_{0}}\Phi(\chi r); \\~\\
  e^{\lambda}=e^{\lambda_{0}}\Psi(\chi r);~~e^{\mu}=e^{\mu_{0}}X(\chi r),
\end{array} \label{2}
\end{eqnarray}
where  $U_0,~e^{\nu_0},~e^{\lambda_0}$ and $e^{\mu_0}$ are solutions for  the case  $\chi= 0$,  and  $V, ~\Phi, ~ \Psi$, and $\psi$ are functions of the one argument $\chi {r}$.

As  $\chi {r}\rightarrow0$, each of these four functions  approaches unity;  i.e., for  $r\rightarrow0$, the solutions  for  the  cases $\chi\neq0$ and $\chi=0$ are  identical.

Let us set \cite{duan4}
\begin{equation}
\mu=-2\ln f(r),\nn
\end{equation}
where  $f(r)$ is an arbitrary function with  no singularities  in the  region $0\leq r\leq +\infty$  and satisfying
the following  conditions:
\begin{eqnarray}
\begin{array}{l}
  ~|f (r)|\leq1;~~[f (r)]_{r=0}=0;~~[f (r)]_{r=\infty}=1; \\~\\
  ~[f '(r)]_{r=0}=\alpha;~~[f '(r)]_{r=\infty}=0; \\~\\
  ~[f ''(r)]_{r=0}=\beta;~~[f ''(r)]_{r=\infty}=0.
\end{array} \label{3}
\end{eqnarray}
Solving the  Einstein equation for  X= 0 and  inserting the solutions for  $U_0, ~ e^{\nu_0}, ~ e^{\lambda_0}$ and $e^{\mu_0}$  into Eq. \eqref{2}, we  arrive  at the  general solution in the form
\begin{eqnarray}
\begin{array}{l}
U_{r}=\frac{G}{\sqrt{A^2+4(kG^2/c^4)}}\times\ln\left[\frac{X+(A/2q)(1+q)}{X-(A/2q)(1-q)}\right]V(\chi r);  \\~\\
e^{\nu}=\left[\frac{X-(A/2q)(1-q)}{X+(A/2q)(1+q)}\right]^q\Phi(\chi r);\\~\\
e^{\lambda}=\frac{r^2}{f^4} \left[1-\frac{r f'(r)}{f(r)}\right]^{q}\frac{1}{X}\left[\frac{X-(A/2q)(1-q)}{X+(A/2q)(1+q)}\right]^q\Psi(\chi r);  \\~\\
e^{\mu}=\frac{X(\chi r)}{f^2(r)},
\end{array} \label{4}
\end{eqnarray}
where  $X$ is the solution to the  algebraic equation
\begin{eqnarray}
 \frac{[X\!+\!(A/2q){(1\!+\!q)}]^{q+1}}{[X\!-\!(A/2q){(1\!-\!q)}]^{q-1}}
    =\frac{r^2}{f^2(r)}, \nonumber\\
q=A\,[A^2\!+\!4(kG^2/c^4)]^{-\frac{1}{2}}<1,
\label{5}
\end{eqnarray}
and $A$ is a  constant of  integration which  can be determined from  the conditions \eqref{3}. From Eq. \eqref{4} and conditions \eqref{3} it is easy to prove that all the components of the metric tensor $g_{ik}$ and  of the potential U are  nonsingular as $r\rightarrow0$.

The self-energy of a particle is calculated using Tolman's formula \cite{tolman}
\begin{eqnarray}
W&=&\int(T^{1}_{1}+T^{2}_{2}+T^{3}_{3}-T^{4}_{4})\sqrt{-g}dV\nonumber \\
 &=& G^2\int^{\infty}_{0}\frac{d(r/f)}{(r/f)X}=\delta.\label{6}
\end{eqnarray}
It  can be shown that with conditions \eqref{3}, the integral in Eq. \eqref{6} is finite. Then the self-mass  of
 the  nucleon will be  $m = G^2\delta/c^2$, and  its classical radius  is $r_0=1/\delta$.  If we  neglect small quantities  of the  order  of  $kG^2/c^4=10^{-68}$, then the  constant $q\approx1$.

In this case the  potential of  the  scalar meson is  of the  form
\begin{equation}\label{7}
U(r)=-\frac{G}{A}~\ln\left[1-A\frac{f(r)}{r}\right]e^{-\chi r}.
\end{equation}
In  view of the  condition $|f(r)|\leq1$, we can expand the potential $U(r)$  in a series for $r >A$:
\begin{eqnarray}\label{8}
U(r)&=&\frac{G}{r}f(r)e^{-\chi r}\left[1+\frac{1}{2}\Big(\frac{A}{r}\Big)f(r) \right. \nonumber \\
 &&\hskip 22mm \left.+\frac{1}{3}\Big(\frac{A}{r}\Big)^2f^2(r)+\cdots \right].\nn
\end{eqnarray}
For $r\gg A$,  the  function $f(r)\approx1$, and we obtain
the  Yukawa  potential
\begin{equation}
U(r)=Ge^{-\chi r}/r.
\end{equation}
Equation \eqref{7}  is a  generalization of the  Yukawa potential.  When  $r\sim A$, the gravitational field
greatly alters the  potential of the  meson field.  From \eqref{7}  and \eqref{3} we  obtain
\begin{equation}
U(0)=(G/A)\ln[1/(1-A\alpha)].
\end{equation}
Thus the  theory we  have  here  developed  is  in a position to give  not only a  finite  nucleon mass, but also a  finite  potential well depth for  nuclear forces.  In  this  lies its attractiveness.

In conclusion I express my  gratitude  to Professor M.  F. Shirokov for  valuable  advice and suggestions during the  performance of this work.

\end{document}